\documentclass[letters,fleqn,usenatbib]{mnras}

\usepackage[T1]{fontenc}
\usepackage{ae,aecompl}
\usepackage[normalem]{ulem}

\usepackage{graphicx}	
\usepackage{amsmath}	
\usepackage{amssymb}
\usepackage{booktabs}

\title[The host galaxy of PKS~1502$+$036]{The host galaxy of the $\gamma$-ray-emitting narrow-line Seyfert 1 galaxy PKS~1502$+$036}

\author[F. D'Ammando et al.]{F. D'Ammando$^{1}$\thanks{E-mail: dammando@ira.inaf.it}, J. A. Acosta-Pulido$^{2,3}$, A. Capetti$^{4}$, R. D. Baldi$^{5}$, M. Orienti$^{1}$ \newauthor C. M. Raiteri$^{4}$, and C. Ramos Almeida$^{2,3}$ \\  
$^{1}$INAF -- Istituto di Radioastronomia, Via Gobetti 101, I-40129 Bologna, Italy \\ 
$^{2}$Instituto de Astrofisica de Canarias, Calle Via Lactea, s/n, E-38205 La Laguna, Tenerife, Spain \\ 
$^{3}$Departamento de Astrofisica, Universidad de La Laguna, E-38205 La Laguna, Tenerife, Spain \\ 
$^{4}$INAF -- Osservatorio Astrofisico di Torino, via Osservatorio 20, I-10025, Pino Torinese, Italy \\
$^{5}$Department of Physics and Astronomy, University of Southampton, Southampton SO17 1BJ, UK }

\date{}

\pubyear{2018}

\begin{document}
\label{firstpage}
\pagerange{\pageref{firstpage}--\pageref{lastpage}}

\maketitle

\begin{abstract}
The detection of $\gamma$-ray emission from narrow-line Seyfert 1 galaxies (NLSy1) has challenged the idea that large black hole (BH) masses ($\ge$10$^8$ M$_{\odot}$) are needed to launch relativistic jets. We present near-infrared imaging data of the $\gamma$-ray-emitting NLSy1 PKS~1502$+$036 obtained with the Very Large Telescope. Its surface brightness profile, extending to $\sim$20 kpc, is well described by the combination of a nuclear component and a bulge with a S\'ersic index $n$ = 3.5, which is indicative of an elliptical galaxy. A circumnuclear structure observed near PKS 1502$+$036 may be the result of galaxy interactions. A BH mass of $\sim$7 $\times$ 10$^{8}$\,M$_{\odot}$ has been estimated by the bulge luminosity. The presence of an additional faint disc component cannot be ruled out with the present data, but this would reduce the BH mass estimate by only $\sim$30 per cent. These results, together with analogous findings obtained for FBQS J1644$+$2619, indicate that the relativistic jets in $\gamma$-ray-emitting NLSy1 are likely produced by massive black holes at the centre of elliptical galaxies.
\end{abstract}


\section{Introduction}

Relativistic jets are the manifestation of the extraordinary amount of energy released by a supermassive black hole (SMBH) in the centre of an active galactic nucleus (AGN). Despite decades of efforts, a consensus about the formation of relativistic jets has been reached neither on the details of the mechanisms nor on the ultimate source of energy: rotational energy from the spinning SMBH or accretion power. Recent studies indicate that jets carry a power larger than that associated with the accretion flow, suggesting that the main source of the jet power should be the SMBH
spin \citep{ghisellini14}, in agreement with general-relativistic magnetohydrodynamics simulations \citep[e.g.][]{tchekhovskoy11}. 

Powerful jets are apparently only associated with the most massive SMBH ($M_{\rm BH}$\,$>$10$^{8}$\,M$_{\odot}$) \citep[e.g.][]{chiaberge11} hosted in elliptical galaxies: only a few powerful radio sources have been observed in disc galaxies, but also in these cases the estimated BH mass is $>$10$^{8}$M$_{\odot}$ \citep[e.g.][]{morganti11,singh15}. This has been interpreted as evidence that efficient jet formation requires rapidly spinning SMBH, which can be obtained through major mergers \citep[e.g.][]{sikora07}. Indeed, major mergers are commonly observed in the elliptical galaxies producing the most powerful jets \citep[e.g.][]{ramos11,chiaberge15}. 

In this context, the detection of variable $\gamma$-ray emission from a dozen radio-loud narrow-line Seyfert 1 galaxies (NLSy1) by the {\em Fermi} satellite \citep[e.g.][]{abdo09,dammando12,dammando15,paliya18}, requiring the presence of relativistic jets in these objects, challenges the theoretical scenarios of jet formation \citep[e.g.][]{boettcher02}. NLSy1 are usually hosted in spiral/disc galaxies \citep[e.g.][]{deo06}, although some of them are associated with early-type S0 galaxies \citep[e.g. Mrk~705 and Mrk~1239;][]{markarian89}. These galaxies are characterized by the presence of pseudo-bulges produced by secular processes \citep[e.g.][]{mathur12}, are usually formed by minor mergers, and have estimated BH masses of, typically, 10$^{6}$--10$^{7}$ M$_\odot$ \citep[e.g.][]{woo02}. This casts doubts on the connection between relativistic jet production, SMBH mass,  and the host galaxy, suggesting that relativistic jets in NLSy1 might be produced by a different mechanism.

However, studies of the properties of the host galaxy of NLSy1 have been done mainly for radio-quiet sources at low ($z$ < 0.1) redshift \citep[e.g.][]{crenshaw03, deo06}. Understanding the nature of the host galaxies of $\gamma$-ray-emitting NLSy1 and estimating their BH mass are of great interest in the context of the models for the formation of relativistic jets \citep[e.g.][]{hopkins05}. 

The morphology of the host galaxy has been investigated only for three such sources, namely 1H~0323$+$342, PKS~2004$-$447, and FBQS~J1644$+$2619. Observations of 1H~0323$+$342 revealed a structure that may be interpreted either as a one-armed spiral galaxy \citep{zhou07} or as a circumnuclear ring produced by a recent merger \citep{anton08, leontavares14}. The analysis of near-infrared (NIR) Very Large Telescope (VLT) observations, suggested a possible pseudo-bulge morphology for the host of PKS~2004$-$447, although the surface brightness distribution of the host is not well constrained by a bulge$+$disc model at large radii \citep{kotilainen16}. A pseudo-bulge morphology of the host galaxy of FBQS~J1644$+$2619 has been also
claimed by \citet{olguin17}, based on Nordic Optical Telescope observations.

We presented deep NIR observations of FBQS~J1644$+$2619 obtained with the Gran Telescopio Canarias. The surface brightness profile of the source is modeled up to 5 arcsec ($\sim$13 kpc) by the combination of a nuclear component and a bulge component with S\'ersic index $n$ = 3.7. This indicates that the relativistic jet in this source is produced in an elliptical galaxy and we could infer a BH mass of $\sim$2 $\times$10$^{8}$\,M$_\odot$ \citep{dammando17}. The debate on the host properties of the $\gamma$-ray-emitting NLSy1 is then still open.

In this Letter, we report an analogous study based on NIR observations of the NLSy1 PKS~1502$+$036 collected using the ISAAC camera at the VLT. PKS~1502$+$036 (also known as J1505$+$0326 and SDSS J150506.47$+$032630.8) is an NLSy1 \citep[e.g.][]{yuan08}, associated with a $\gamma$-ray source \citep{dammando13}, at redshift $z$ = 0.408 \citep{hewett10}, where 1 arcsec corresponds to 5.26 kpc. The manuscript is organized as follows. The VLT data analysis and results are presented in Section 2. In Section 3, we discuss the host galaxy morphology and the BH mass estimate, comparing our results to those obtained for the other $\gamma$-ray-emitting NLSy1 and summarize our conclusions.

\section{Data Analysis}

\begin{figure*}
\centering
\includegraphics[width=17.7cm]{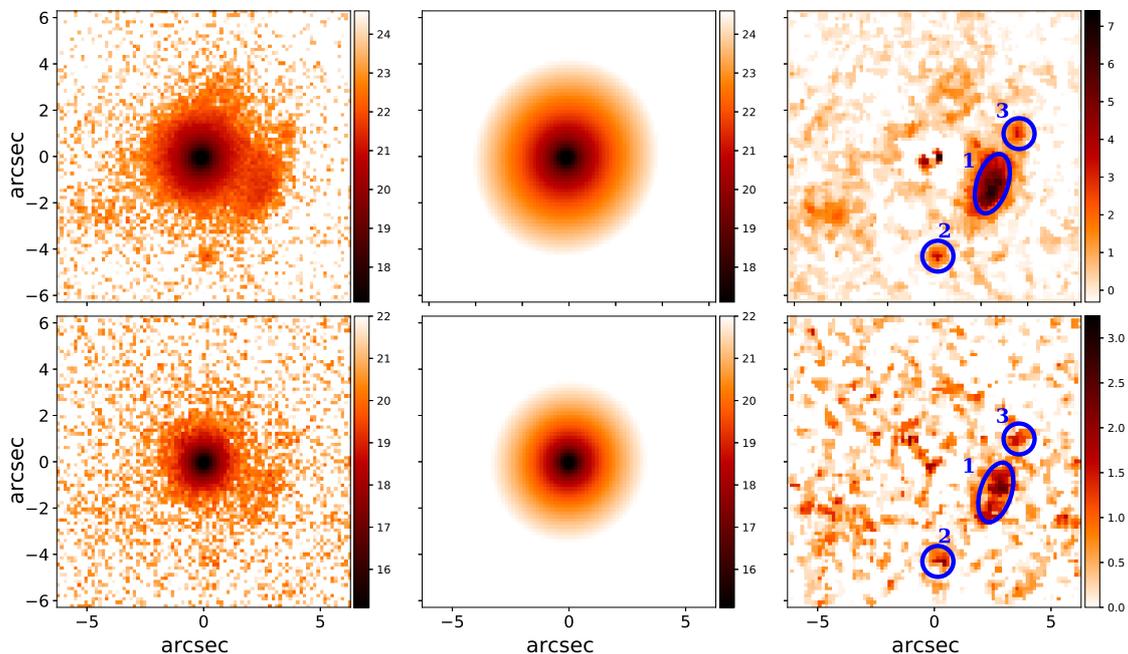}
\caption{Left: central 13 $\times$ 13 arcsec$^{2}$ (68 $\times$ 68 kpc$^{2}$) of the images in the $J$ and $K_s$ band of PKS~1502$+$036, top and bottom, respectively. Center: \texttt{GALFIT} models using a S\'ersic profile combined with a nuclear PSF. Right: residual images after subtracting the model; the blue ellipses mark a structure for which we extract photometry. Colour bars are in mag arcsec$^{-2}$ (left-hand and centre panels) and ADU/RMS (right-hand panels). In all panels, north is up and east is right.}
\label{fig:galfmod}
\end{figure*}

\begin{figure*}
\includegraphics[width=8.5cm]{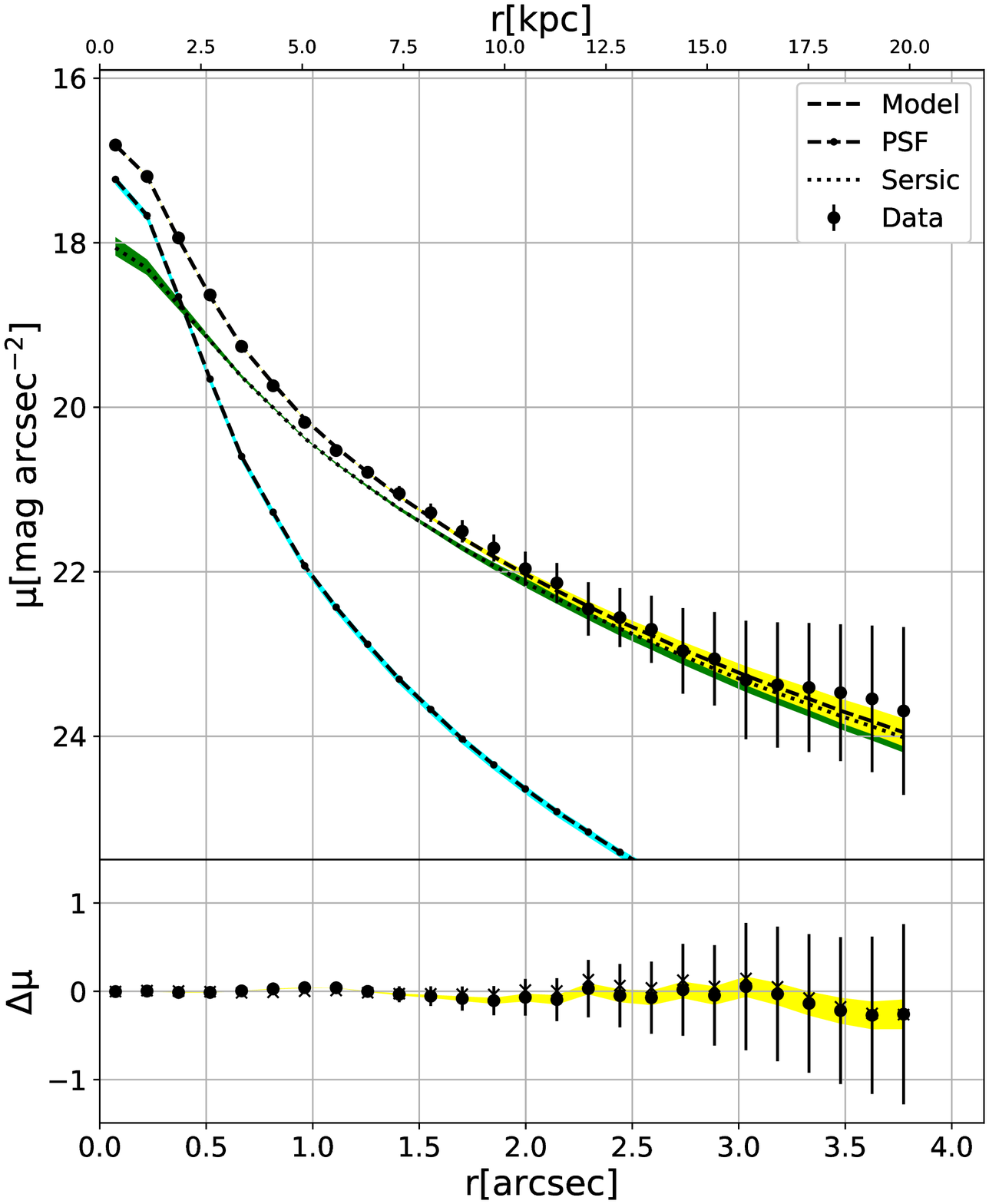}
\includegraphics[width=8.5cm]{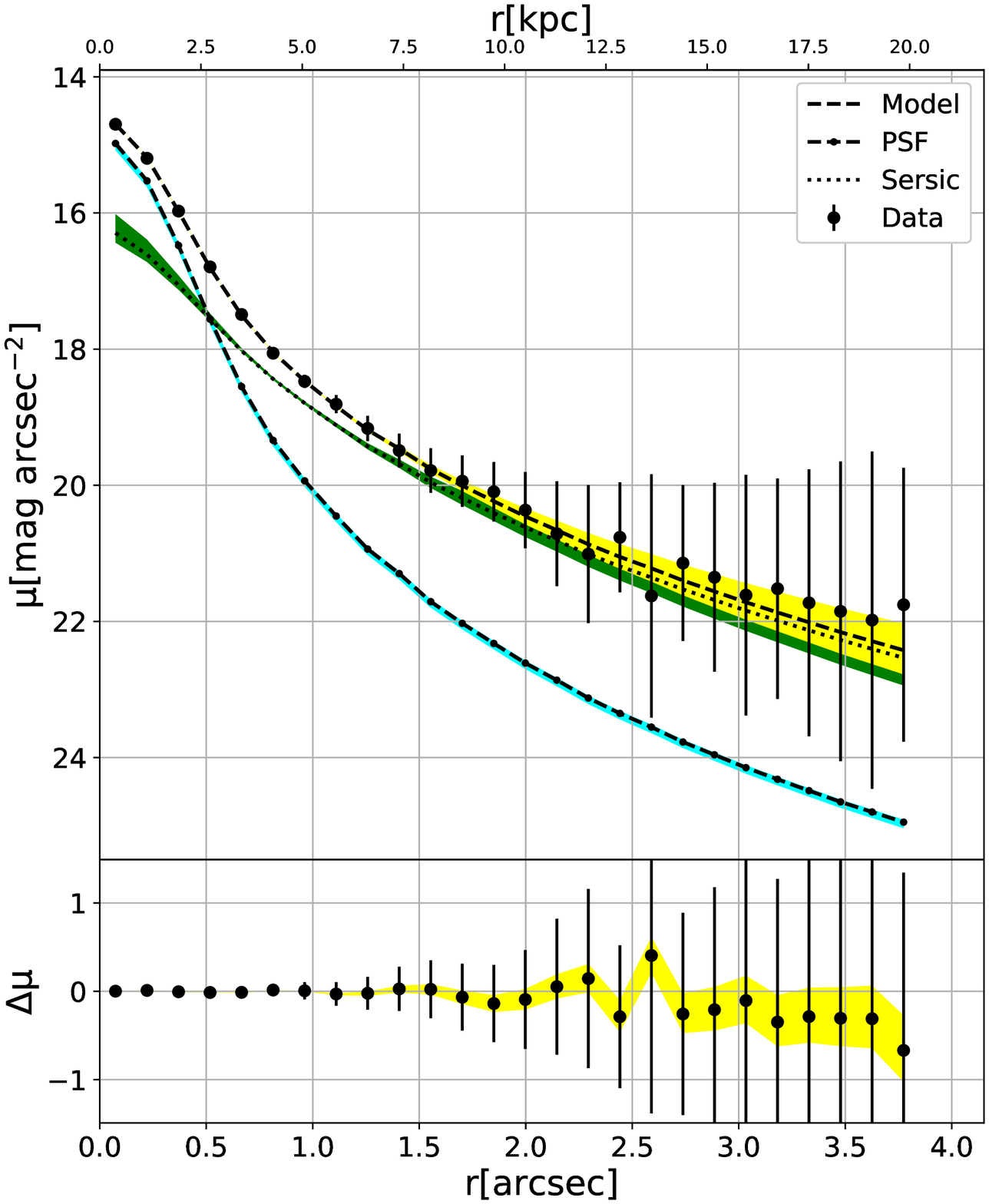}
\caption{1D surface brightness decomposition in the $J$ (left) and $K_s$ (right) bands obtained by using Star 1 as PSF template. The observed profiles are the black dots, the nuclear PSF is the dot-dashed light blue curve, the bulge component is reproduced with a S\'ersic model (green dot line). In the bottom panels we show the residuals. In the $J$-band residual panel crosses represent bulge+disc component.}
\label{fig:2dmod}
\end{figure*} 

\subsection{Observations}

VLT/ISAAC \citep{moorwood99} $J$- and $K_s$-band images were retrieved from ESO scientific archive [programme 290.B-5045(A)]. ISAAC is equipped with a $1024\times1024$ pixels Hawaii Rockwell array, with a pixel scale of 0.147 arcsec. The data in $J$ and $K_s$ bands were obtained under excellent seeing conditions (FWHM $\simeq$~0.40 arcsec and FWHM $\simeq$~ 0.37 arcsec for the $J$ and $K_s$ band, respectively), on 2013 March 18 and April 14, respectively. The images were obtained following a random jitter pattern of 23 and 14 positions in the $J$ and $K_s$ bands, respectively, and with a mean separation of 5 arcsec. At each position, three repetitions of 40 sec were taken in the $J$ band and 6 repetitions of 10 sec in the $K_s$ band, providing a total integration time of 46 min for the $J$ band and 14 min for the $K_s$ band. The data reduction was performed using the jitter task within {\sl eclipse} \citep{Devillard99}, including flat-fielding, background subtraction, registration, and combination of individual frames. In addition we have introduced a correction for the reset anomaly effect, commonly present in Hawaii-I arrays. This effect shows up as a brightening of the bottom rows at each of the four array quadrants, which can be successfully removed by fitting two surfaces with a moderate vertical gradient. The photometric zero-points were taken from the ESO Quality Control web pages.\footnote{{\tt https://www.eso.org/observing/dfo/quality/index\textunderscore isaac.html}} The S/N ratio of the target measured from the peak is $\sim$1000 in $J$ band and $\sim$500 in $K_s$ band.

\subsection{Host galaxy structure}

Fig.~\ref{fig:galfmod} (left-hand panels) shows the central 13 $\times$ 13 arcsec$^2$ (68 $\times$ 68 kpc$^2$) of the $J$-\,(top) and $K$-\,(bottom) band images. Besides the NLSy1 host galaxy several other sources are seen, the brightest of which is an extended region centered $\sim$3 arcsec NE: it was masked before modeling. We model the host galaxy images with the 2D surface brightness model fitting code \texttt{GALFIT} \citep{Peng02}, including a bulge component, characterized by a S\'ersic profile, and a point spread function (PSF) to fit the unresolved nuclear component. The normally sampled PSF was obtained independently from the brightest star in the field (Star 1) and from an average profile using three stars (Star 1: $m_J$ = 16.651, $m_K$ = 15.977; Star 2: $m_J$ = 17.525, $m_K$ = 17.137; Star 3: $m_J$ = 17.862, $m_K$ = 17.136; see Fig.~A\ref{appendice} in the Appendix),  which were fit with a Moffat function plus an additional Gaussian for the wings.

All parameters of the S\'ersic profile and the scaling factor of the PSF were allowed to vary. The resulting models are shown in the central panels of Fig.~\ref{fig:galfmod}, the residuals are given in the right-hand panels. The best-fitting values are provided in Table~\ref{ta:galfpar}. We estimated the parameter uncertainties by varying the sky value by $\pm 1 \sigma$ rms value.

1D brightness profiles of PKS~1502$+$036 model, obtained by using Star 1 as PSF template, are shown in Fig.~\ref{fig:2dmod}. The profiles extend to a radius $r \sim$3.7 arsec ($\sim$20 kpc). The nuclear contribution at 1 arcsec is $\sim$2 magnitudes fainter than the galactic one and this permits a detailed study of the structural parameters of the host.  

In the following we adopt the value derived from Star 1, the brightest reference star. The best-fitting S\'ersic index is $n=3.5$, which is a good
description of an elliptical galaxy \citep{blanton03}. From the 2D modeling we obtain an ellipticity of $\epsilon = 1 - b/a = 0.09$ and 0.05 in $J$ and $K_s$ filter, respectively, indicating that the host of PKS~1502$+$036 is an E1-type galaxy.

The absolute magnitudes are computed using a distance module of 41.67, Galactic extinction corrections of $A_J=0.033$ and $A_K=0.017$, and
$K$-corrections of 0.0 and 0.58 mag in the $J$ and $K_s$ band, respectively, adopting the values provided by \citet{Chilingarian10} for luminous red galaxies. The resulting color of the host is $J-K_s=0.91$ in good agreement with the value measured in local elliptical galaxies, $J-K_s=0.87$
\citep{mannucci01}. From the 2D modelling and the 1D profile, we conclude that a bulge component is adequate for describing the host galaxy structure in our NIR images up to a radius of $\sim$3.7 arcsec ($\sim$20 kpc).

\begin{table*}
\caption{Photometric and structural parameters of the PKS~1502$+$036 host galaxy.} 
\label{ta:galfpar}
\resizebox{\textwidth}{!}{\begin{tabular}{lccc|cccccccccc}
\hline
Band/Star & \multicolumn{2}{c}{PSF} & \multicolumn{5}{c}{Bulge--S\'ersic model} & \multicolumn{3}{c}{Disc} & \multicolumn{1}{c}{$\chi^{2}$$_{\nu}$} \\
 \cmidrule(l{10pt}r{5pt}){2-3} \cmidrule(l{10pt}r{5pt}){4-8} \cmidrule(l{10pt}r{5pt}){9-11} \\
 &       & FWHM             &     &     &                      &       &  &   & & & \\
 & mag   & (arcsec)    & mag & $n$ & $R_e$ (arcsec / kpc) & $b/a$ & <$\mu_e$>  &  mag  & $R_s$ (arcsec / kpc) & $b/a$ &  \\
\hline
PSF+bulge & \multicolumn{10}{c}{ } \\
$J$~/ Star 1  & 17.87$\pm$0.04 &  0.40  & 17.29$\pm$0.10 & 3.5$\pm$ 0.8 &  0.84$\pm$0.06/4.4 & 0.91$\pm$0.01 & 18.90$\pm$0.15 & ... & ... & ... &  1.152\\
$J$~/ avg PSF & 17.76$\pm$0.03 &  0.40  & 17.36$\pm$0.08 & 3.3$\pm$ 0.3 &  0.81$\pm$0.02/4.3 & 0.91$\pm$0.01 & 18.87$\pm$0.12 & ... & ... & ... &  1.151\\
$K_s$~/ Star 1  & 15.93$\pm$0.05 &  0.37  & 15.76$\pm$ 0.22 & 3.5$\pm$ 1.7 &  0.76$\pm$0.16/4.0 & 0.95$\pm$0.02 &  17.16$\pm$0.24 & ... & ... & ... &  1.085 \\
$K_s$~/ avg PSF & 15.93$\pm$0.04 &  0.37  & 15.75$\pm$ 0.21 & 3.2$\pm$ 1.3 &  0.93$\pm$0.24/4.9 & 0.97$\pm$0.02 &  17.59$\pm$0.23 & ... & ... & ... &  1.082 \\
\hline
PSF+bulge+disc & \multicolumn{10}{c}{ } \\
$J$~/ Star 1  & 18.20$\pm$0.10 &  0.40  & 17.66$\pm$0.04 & 4.0 & 0.26$\pm$0.07/1.4 & 0.85$\pm$0.01 &  16.69$\pm$0.05 & 18.25$\pm$0.07 & 1.02$\pm$0.14/5.4  & 0.98$\pm$0.01 &  1.138\\
$J$~/ avg PSF & 18.03$\pm$0.06 &  0.40  & 17.74$\pm$0.14 & 4.0 & 0.32$\pm$0.04/1.7 & 0.87$\pm$0.01 &  17.24$\pm$0.15 & 18.30$\pm$0.15 & 1.01$\pm$0.13/5.3 & 0.98$\pm$0.01 &  1.129\\
\hline
\end{tabular}}
\end{table*}

The BH mass was computed using the relationship between the NIR bulge luminosity and the BH mass provided by \citet{marconi03} for the sub-group of galaxies with a secure BH mass estimate (i.e. group 1; see their table 2). The resulting BH mass estimated from the $J$ and $K_s$ values is
7.3$\times$10$^{8}$\,M$_{\odot}$ and 6.8$\times$10$^{8}$\,M$_{\odot}$, respectively. Based on the dispersion in the \citet{marconi03} $L$--$M_{\rm BH}$ relation, we estimate a factor of $\sim$2 uncertainty on $M_{\rm BH}$.

For the $J$ band, in which the S/N ratio is higher, we consider also the effects of forcing the presence of a disc contribution: in this case an exponential disc profile was added to the bulge component (with index fixed to $n=4$). The fit is just slightly better with respect to considering only a bulge component (Table~\ref{ta:galfpar}). The inclusion of the disc reduces the total flux in the bulge by 0.32 mag. The corresponding BH mass is 4.8$\times$10$^{8}$\,M$_{\odot}$, consistent with the value derived with just a bulge component within the uncertainties. Replacing the exponential disc profile with an edge-on disc profile does not improve the fit.

The residual images obtained after 2D \texttt{GALFIT} modelling (Fig.~\ref{fig:galfmod}, right-hand panel) indicate the presence of a structure at a distance of 3 arcsec ($\sim$16 kpc). This structure appears as bright spots ranging from PA $\sim$270 to PA $\sim$370 (marked as Region 1, 2, and 3 in Fig.~\ref{fig:galfmod}). A dimmer and more extended bar-like structure, going from ESE to N, can be observed at the other side of the galaxy centre. In order to get some insight about the origin of this structure we extract photometry from the bright region, Region 1, obtaining: $m_J$ = 21.04 $\pm$ 0.02, $m_K$ = 19.73 $\pm$ 0.07. The resulting color index, $J-K_s$ = 0.70 $\pm$ 0.09, is similar to the value obtained for the bulge component, suggesting that this region lies at a similar redshift of PKS 1502$+$036.

\section{Discussion and conclusions}

The excellent quality of the VLT images and the broad extent of the brightness profile, expanding to $\sim$20 kpc, enable us to study in detail the properties of the host of the $\gamma$-ray-emitting NLSy1 PKS~1502$+$036. Our analysis indicates that the host of the source is a luminous ($M_J$ = $-$24.38) elliptical galaxy. The presence of a faint disc component cannot be ruled out with the present data.

The structure appearing in the residual image obtained after 2D \texttt{GALFIT} modelling and formed by three regions offset by 3 arcsec with respect to the nucleus may interpreted as a broken ring resulting from an interaction with another galaxy. It could correspond either to a small companion being destroyed spiraling towards the massive nucleus, or the passage of a spherical companion through a disc galaxy, similar to those forming ring galaxies \citep{fiacconi12}. Given the absence of a prominent disc, we favour the first scenario. Our results resembles those reported by \citet{leontavares14} for 1H 0323$+$342. 

We have also explored the environment of the host galaxy of PKS 1502$+$036 by looking at the SDSS data of a few nearby targets for which spectroscopy is available. Among them there is a bright galaxy ($J$ = 16.8 mag; $K_s$ = 15.35 mag) at about 80 arcsec ($\sim$420 kpc) N from PKS 1502$+$036. Its spectroscopic redshift, $z$ = 0.409, is nearly identical to that of our NLSy1, and the galaxy exhibits spectrum characteristics of early type galaxies. In order to investigate if other nearby galaxies at a similar redshift are present, we retrieve all photometric redshift determinations from the SDSS Sky Server in a field of view of 7 arcmin around PKS 1502$+$036 (see Fig.~A\ref{appendice2} in the Appendix). We find an excess of galaxies showing similar photometric redshift values to that of the mentioned early type galaxy, which may indicate the presence of a group of galaxies at such position. This finding provides additional support to our hypothesis that the circumnuclear structure described before may be the result of galaxy interactions. 

We obtained a SMBH mass of $\sim$7$\times$10$^{8}$\,M$_{\odot}$ from the NIR bulge luminosity of the host galaxy of PKS 1502$+$036. From its optical spectrum and the broad-line region (BLR) radius--luminosity relation by \citet{kaspi05}, a virial mass $M_{\rm BH}$ = 4$\times$10$^6$\,M$_\odot$ has been estimated by \citet{yuan08}. By modelling the optical--UV data with a Shakura \& Sunyaev accretion disc spectrum, \citet{calderone13} found instead $M_{\rm BH}$ = 3$\times$10$^{8}$\,M$_{\odot}$.

Conflicting results were also obtained for FBQS J1644$+$2619. We found that this NLSy1 is hosted by a luminous E1 galaxy with a SMBH mass of 2.1$\times$10$^8$\,M$_\odot$ \citep{dammando17}, again significantly larger than the virial estimate (0.8--1.4$\times$10$^7$\,M$_\odot$, \citealt{yuan08,foschini15}), but compatible with that obtained from the accretion disc emission ($M_{\rm BH}$ = 1.6$\times$10$^{8}$\,M$_{\odot}$; \citealt{calderone13}.).

Similarly, for 1H~0323$+$342 values in the range (1.5--2.2)$\times$10$^7$\,M$_\odot$ were estimated from NIR and optical spectroscopy \citep{landt17}, while values of (1.6--4.0)$\times$10$^8$\,M$_\odot$ were obtained by the NIR bulge luminosity \citep{leontavares14}. 

It appears that the SMBH masses estimated with the virial method in $\gamma$-ray-emitting NLSy1 are systematically and significantly smaller that those derived from other techniques. This discrepancy may be due to the radiation pressure from ionizing photons acting on the BLR clouds \citep{marconi08} or to projection effects on a disc-like BLR \citep{decarli08, baldi16}. The effect of flattening of the BLR on the virial mass estimates may be larger in the $\gamma$-ray-emitting NLSy1, for which the observer's angle of view should be small ($\theta$ $\sim$ 5--10 deg), as suggested by their blazar-like behaviour \citep[see e.g.][and the reference therein]{dammando16}. 

These findings represent increasing evidence that the hosts of $\gamma$-ray-emitting (and radio-loud) NLSy1 differ from those of radio-quiet NLSy1, generally spirals with low-BH masses. In the case of PKS~1502$+$036, as well as FBQS~J1644$+$2619 \citep{dammando17}, the host is a E1 galaxy and the BH mass is 2--7$\times$10$^{8}$\,M$_\odot$, in agreement with what is observed in radio-loud AGN. These observational results confirm that a massive SMBH is a key ingredient for launching a relativistic jet that, among NLSy1, only those hosted in  massive elliptical galaxies are able to produce. This is related to the fact that jet power arises from both mass (thus the accretion) and spin of the BH, and mainly by means of the major mergers occurred in elliptical galaxies the SMBH can be significantly spinned-up. This is clearly a key issue in the context of our understanding of the production of powerful relativistic jets in radio-loud AGN.

The number of $\gamma$-ray-emitting NLSy1 is still rather limited and the literature shows conflicting results. Further high-spatial resolution
observations of their host galaxies are needed to establish whether these sources are hosted in spiral or elliptical galaxies and, even more important in the context of the production of a relativistic jet, to measure their BH mass.

\section*{Acknowledgements}

Based on observations made with ESO Telescopes at the La Silla Paranal Observatory under programme 290.B-5045.

\label{lastpage}

\appendix

\onecolumn

\begin{figure}
\centering
\includegraphics[width=9.5cm]{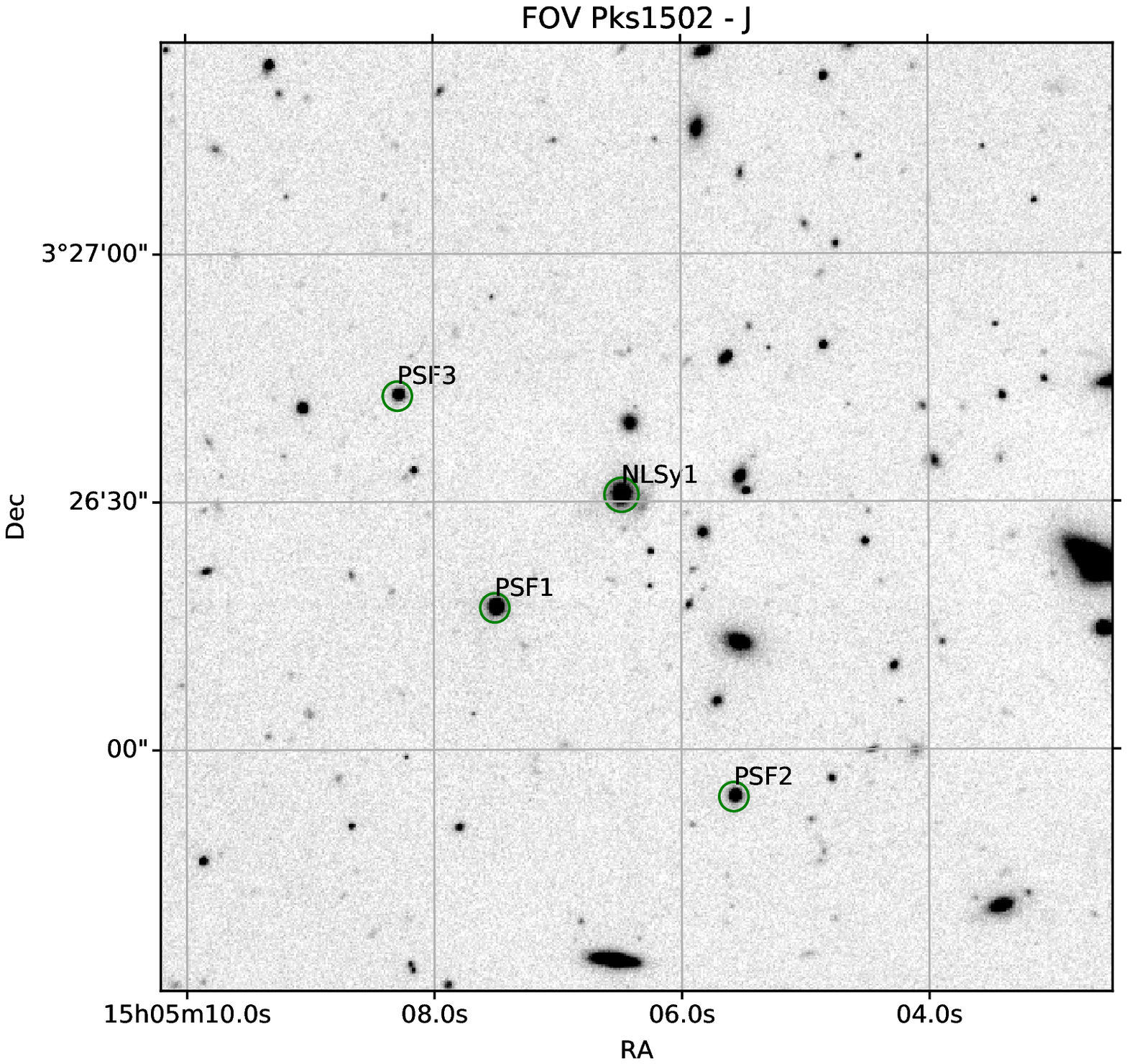}
\caption{ISAAC image (150 $\times$ 105 arcsec$^2$, i.e. 789 $\times$ 552 kpc) in the $J$ band. We marked the location of the target and of the three stars used for producing the PSF templates.}
\label{appendice}
\end{figure}

\begin{figure}
\centering
\includegraphics[width=10.0cm]{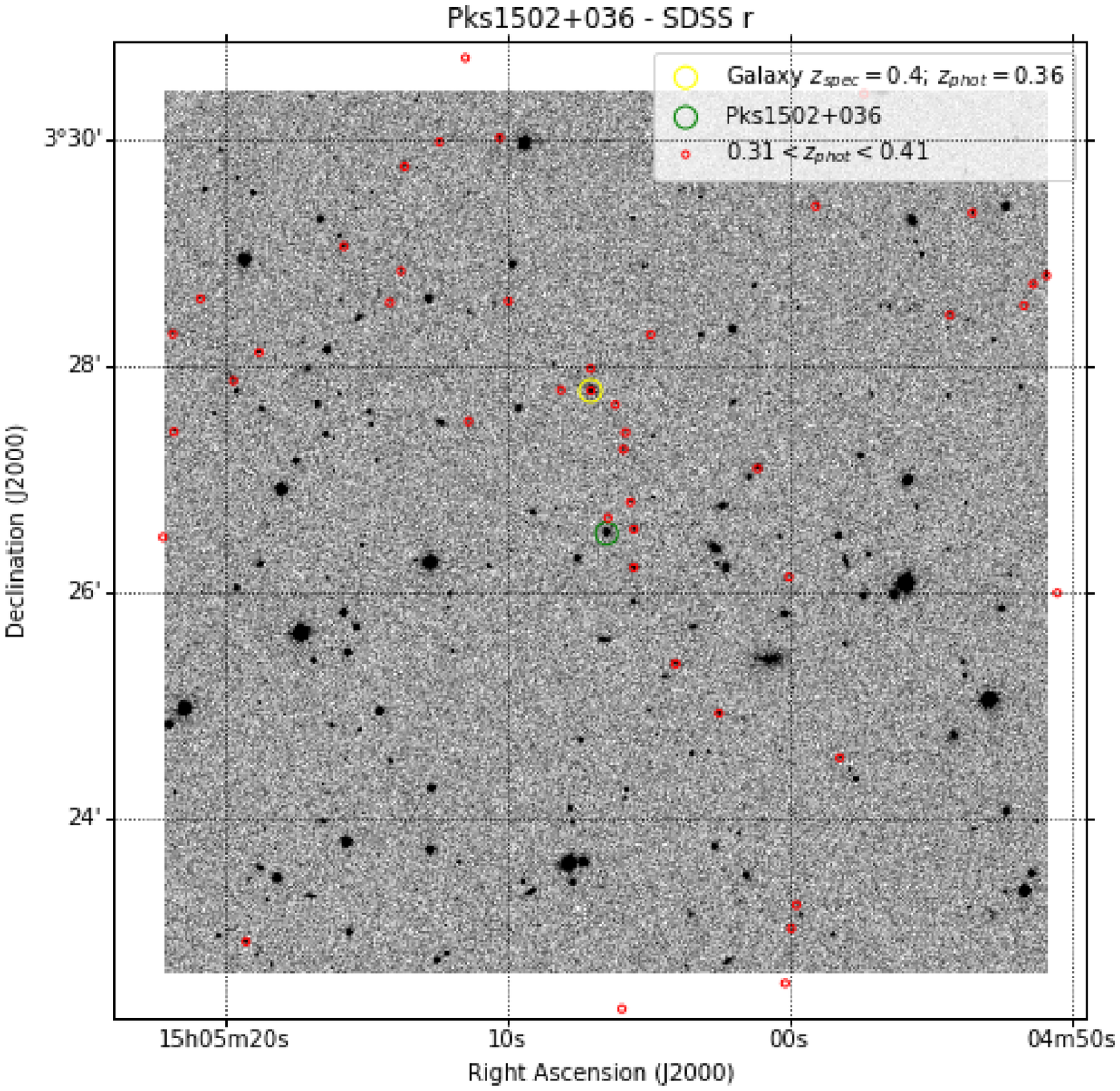}
\caption{SDSS $r$-band image in a field of view of 7 arcmin around PKS 1502$+$036, identified by a green circle. Galaxies with similar photometric
 redshift of our target are highlighted with red circles, and a bright galaxy almost at the same redshift of our target with a yellow circle. }
\label{appendice2}
\end{figure}

\end{document}